\documentstyle[11pt,newpasp,twoside,epsf]{article}
\def\edcomment#1{\iffalse\marginpar{\raggedright\sl#1\/}\else\relax\fi}
\reversemarginpar

\begin{document}
\title{The CNO problem in magnetic CVs}
 \author{Jean-Marc Bonnet-Bidaud}
\affil{Service d'Astrophysique, DSM/DAPNIA/SAp, CEA Saclay, 91191 Gif-sur-Yvette, France}
\author{Martine Mouchet}
\affil{Observatoire de Paris, 92190 Meudon, France}

\begin{abstract}
Some polars like BY Cam are characterized by unusual CNO line ratios compared to other 
polars and non-solar abundances have been suggested to explain this anomaly.
We present here a first attempt to constrain the elemental abundances in these systems 
by applying a specific ionisation model combined with a geometrical 
description of the accretion column where these lines are thought to be formed. 
The line luminosities have been computed using the CLOUDY plasma code for different 
ionisation spectra and column extension. We show here selected results and compare to 
the values observed in "peculiar" magnetic CVs.
The model applied to BY Cam confirms that ionization models with solar abundances 
fail to reproduce the observed line intensity ratios. Assuming the model to be valid,
the induced best abundances imply an overabundance of N (x25), underabundance of C (:8)
and nearly solar O (:2), in line with CNO reprocessing.
\end{abstract}

\section{Introduction}
Magnetic Cataclysmic Variables (MCVs are 
close binary systems where a magnetic white dwarf accretes matter from a low-mass companion 
via an accretion column. 
The broad emission lines observed in the optical, mostly hydrogen and helium, are thought to arise 
from irradiation of this column by the X-ray flux emerging from a shock above the heated polar 
cap of the white dwarf. 
Ultraviolet observations with the IUE satellite have also revealed among MCVs the existence of strong 
resonance lines of silicium SiIV($\lambda$1397), nitrogen NV($\lambda$1240) 
and carbon CIV($\lambda$1549). Most sources were found to have line intensity ratios NV/CIV and 
SiIV/CIV similar to all other CVs (see Mauche et al. 1997, for a review). 

A new situation arised however when the source 
BY Cam was observed with IUE, shortly after its discovery as a polar, 
on the basis of its optical polarisation by Remillard et al. (1986). 
It was shown to display an impressive NV($\lambda$1240) line with an intensity greater 
than CIV($\lambda$1549), an "inverted ratio" with respect to other MCVs 
(Bonnet-Bidaud \& Mouchet 1987, BM87)). 
It was noted at that time that this unusual feature could be due to non-solar abundances
resulting from an unnoticed nova event or from a chemical evolution of the secondary 
(BM87, Mouchet et al. 1990). 
Shortly after, the discovery that BY Cam is  slightly desynchronised 
(Silber et al. 1992) and the identification of a known historical nova, V1500 Cygni, 
as a polar which is also desynchronised (Schmidt \& Stockman 1991), gave 
further arguments to the nova hypothesis. 
However HST observations of more recent MCVs have gradually changed this picture since at least 
two synchronous polars, V1309 Ori (Szkody \& Silber 1996, Schmidt 
\& Stockman 2001) and MN Hya (Schmidt \& Stockman 2001), have now been found to also 
show a large NV/CIV ratio, though less extreme than in BY Cam. 
Figure 1 summarises the present situation. Most CVs nicely cluster around
similar values, including MCVs (shown by typical values observed for AM Her), while an order 
magnitude difference is observed for BY Cam, V1309 Ori and MN Hya. Also noticeable are the
extreme values observed for AE Aqr.

The number of anomalous sources now points to what can be called the "CNO problem" among MCVs.
The explanation of these discrepant line ratios is still pending.
An abundance effect is an obvious possibility but as the lines are most probably 
formed by photoionisation, differences may also arise from specific ionisation conditions.
We present here the results of line computations for BY Cam which show that only significant non-solar
abundances can reproduce the observed line ratios.

%=====================================================
\begin{figure}
\plotfiddle{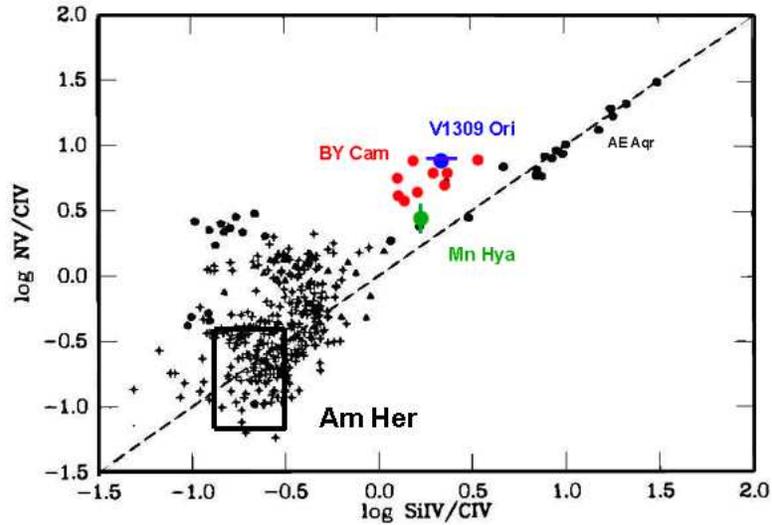}{7cm}{0}{50}{50}{-144}{-100}
\caption{The CNO problem among MCVs. Line flux ratios are shown in logarithmic scale
for all magnetic and non-magnetic CVs (crosses) and AE Aqr (black dots), BY Cam (grey dots), 
V1309 Ori (dot/horizontal line), MN Hya ((dot/vertical line). The box shows the
range of values observed for AM Her (adapted from Mauche et al., 1997)}
\end{figure}
%=====================================================

\section{Model of the accretion column}
To compute the line flux, we adopt an approach similar to the work of Stockman \& Schmidt (1996) by
coupling a photoionized code with a simple description of the accretion column.
Following Langer et al. (1982), we model the accretion column 
with a variable cross-section according to the dipole geometry and  free-fall velocities along the column,
yielding a density varying with the distance $r$ to the white dwarf, 
as \,$n \sim n_0\,(r/R_{wd})^{-2.5}$. 
The density, n$_0$, at the basis of the column, is computed from the accretion luminosity L$_{X}$
as n$_0$ =$1.8\times 10^{16}$ cm$^{-3}$\, (L$_{X}$/$10^{34}$ erg\,s$^{-1}$), assuming a typical value  
of 10$^{16}$\,cm$^2$ for the polar cap surface, and a 0.8 M$_{\sun}$  white dwarf.
The accretion column was approximated by a succession of different slabs 
of constant densities with a maximum extension chosen so that
the density drops by a factor 4 from one slab to the next, up to a maximum 
extension of 50 times the white dwarf radius, i.e. about 1/4 of the 
Roche lobe radius (Mouchet et al. 1997). The lateral extension of each region
was assumed to follow the dipole geometry with a size varying as (r/R$_{wd}$)$^{3/2}$.

%=====================================================
\begin{figure}
\plotfiddle{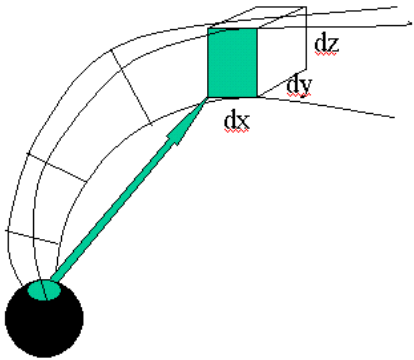}{6cm}{0}{50}{50}{-144}{-100}
\caption{The geometrical model of the accretion column (non-scaled schematic view). 
The cross section is taken to vary according to
the dipole law and vertical constant density slabs are defined with direct sideways X-ray 
illumination from the post-shock region (see text).}
\end{figure}
%=====================================================

Line intensities are computed using the photoionisation code CLOUDY (Ferland et al. 1998).
Contribution from the different constant density slabs are added, considering only direct unobscured
sideways illumination. Taking into account of the bending of the column, this is of course 
only strictly true for the highest parts of the column. 
It was verified however that, in all cases, the lowest slabs of higher 
densities do not contribute significantly to the total flux of the considered lines 
(see below). In the same way, it was checked that for a minimum height above the white dwarf
$h/R_{wd} ~\geq ~5$, the density and column density are respectively lower than $10^{13}$ cm$^{-3}$ and
$10^{23}$ cm$^{-2}$, in the range of acceptable values for CLOUDY.  

\section{Line intensities of BY Cam}
The luminosities of the three resonance lines CIV ($\lambda$1549), 
NV ($\lambda$1240) and OVI ($\lambda$1035) were computed for BY Cam with this model, 
for different ionizing spectra and element abundances.
Figure 3 shows the result corresponding to the observed "normal" spectrum (model M1),
defined as the sum of a 20\,keV bremsstrahlung, as observed in the  RXTE observations 
(Mukai, private communication) and of a
50\,eV blackbody with a bolometric luminosity  of 0.1 times the hard
X-ray luminosity (Ramsay et al. 1994) with solar abundances.
The ionic fractions and line luminosities,
after a sharp increase close to the white dwarf, are 
nearly constant in the upper part of the column, 
CIV being the dominant species over NV and OVI. 
This yields N/C and O/C line ratios of respectively 0.16 and 0.19, compared to the observed 
values of 5.70 and 1.14 (see Table 1).
As the ionisation structure is particularly dependent on the soft part of the ionizing spectrum, 
we also investigate the possibility of an unseen "soft component". 
For a neutral H interstellar column density of $2\times 10^{20}$ cm$^{-2}$, it is found that
a 10\,eV blackbody component  with a luminosity similar to that of the 
bremsstrahlung component, could remain undetected and still be compatible with the flux upper 
limit at 100\,\AA\, of $6\times 10^{-15}$ erg\,cm$^{-2}$\,s$^{-1}$\AA$^{-1}$ 
derived from the EUVE observations (Howell, private communication). 
Corresponding results for this spectrum (model M2)
are shown in Fig. 4 (top) and values listed in Table 1. The N contribution is increased and C 
decreased but even for this extreme spectrum, the predicted line ratios still fall short 
of the observed ones.

Assuming this simplified model of the accretion model to be valid, a way is provided to measure the
elemental abundances. As the source is somewhat variable, no exact fit was attempted but the observed
mean values can be reproduced by keeping the standard spectrum and significantly altering the abundances 
with an overabundance of N (x25), underabundance of C (:8) and nearly solar value for O (:2) 
(model M3, Fig. 4 (bottom) and Table 1). Note that these values are only indicative and that slightly 
different other combinations can be found which match the observed values.

%=====================================================
\begin{figure}
\plotfiddle{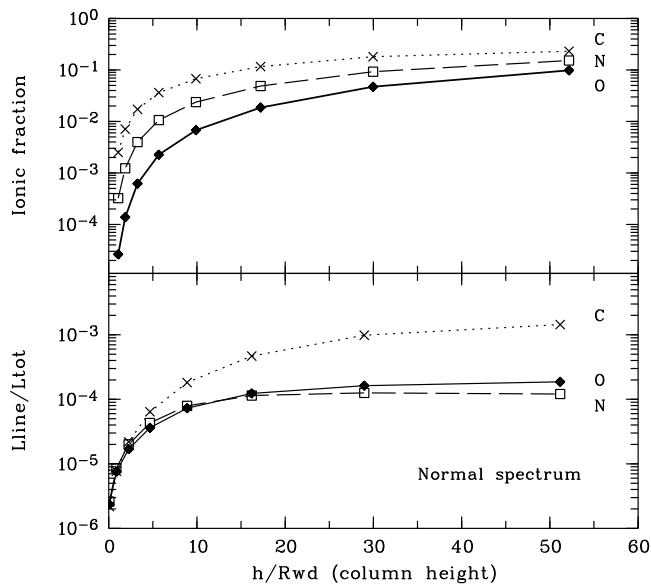}{8cm}{-90}{48}{48}{-144}{260}
\caption{CNO lithium-like ionic fractions (top) and line luminosities (bottom) along 
the accretion column for the BY Cam "normal spectrum" and solar abundances (model M1). 
Line luminosities are expressed in fraction of the total ionizing X-ray luminosity}
\end{figure}
%=====================================================

%=====================================================
\begin{figure}
\plotfiddle{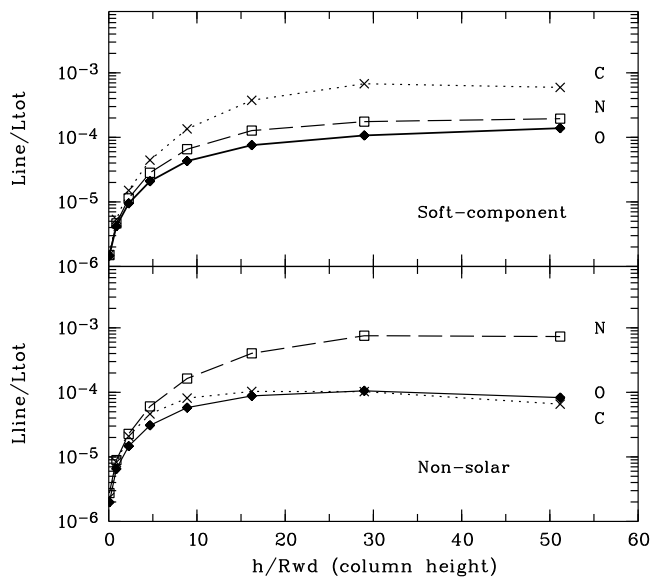}{8cm}{-90}{48}{48}{-144}{260}
\caption{CNO line luminosities along the accretion column for a "soft component" (model 2, top)
and a normal spectrum with non-solar abundances with N (x25), C (:8)
and O (:2) (model 3, bottom)}
\end{figure}
%=====================================================

\begin{table}
\caption{BY Cam CNO line ratios}
\begin{flushleft}
\begin{tabular}{cccccc}
\hline
\multicolumn{1}{c}{Ratio}    & \multicolumn{1}{c}{Observed*} & 
\multicolumn{1}{c}{Dereddened*}  & \multicolumn{1}{c}{Model M1} &
\multicolumn{1}{c}{Model M2}    & \multicolumn{1}{c}{Model M3}  \\
\multicolumn{1}{c}{ }    & \multicolumn{1}{c}{ } & 
\multicolumn{1}{c}{ }  & \multicolumn{1}{c}{Normal} &
\multicolumn{1}{c}{Soft}    & \multicolumn{1}{c}{Non-solar}  \\
\hline
N/C	&  5.30	& 5.70 	& 0.16	& 0.23	& 5.00  \\
O/C   &  0.92	& 1.14	& 0.19	& 0.22	& 0.90  \\
O/N   &  0.17	& 0.20	& 1.18	& 0.66	& 0.18  \\
\hline
\end{tabular}
\end{flushleft}
* values from Mouchet et al. (2003) and these proceedings \\
\end{table}

\section{Discussion}
FUSE, HUT and Orfeus observations, combined to previous ultraviolet results, now provide 
a complete view of the CNO lithium-like lines produced in MCVs.
Anomalous line ratios are found for at least three sources. Though some variability may be invoked,
at least in the case of BY Cam, the overall feature is stable around values an order of 
magnitude different from other MCVs. Photoionisation models of the accretion column, such as presented
here, fail to reproduce these ratios, even by strongly varying the ionisation conditions, inside the range 
compatible with the observed spectrum. Some care should be taken in interpreting the results since 
important simplifications have to be introduced, in particular by considering an homogeneous 
accretion column. The case of a clumpy flow with blob accretion have to be further considered with the
possibility of a dominant collisional ionisation.  

%=====================================================
\begin{figure}
\plotfiddle{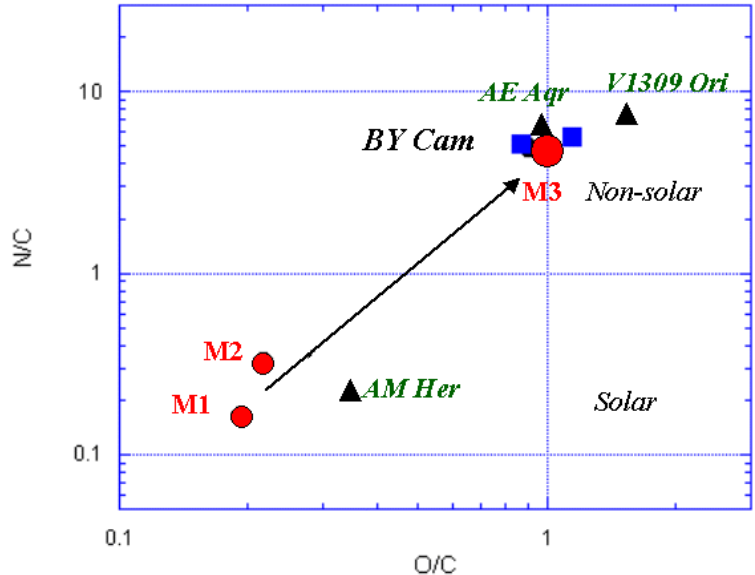}{7cm}{0}{50}{50}{-144}{-100}
\caption{The CNO ratios for magnetic CV's. Observed values are shown by squares for
BY Cam (observed and dereddened) and triangles for AM Her, V1309 Ori and AE Aqr (from Mouchet et al. 2002). 
Grey dots shows theoretical values for BY Cam, assuming solar abundances with 
a "normal" (model M1, 50eV)" or "soft" (model M2, 10eV) spectrum (lower left) and 
non-solar (model M3) values (upper right)}
\end{figure}
%=====================================================

In view of the large variations in the line ratios, a change in the abundances as deduced in the present
work is however strongly suggested. For BY Cam, the observed N/C and O/C ratios points toward a 
CNO redistribution in which carbon is depleted in favour of nitrogen while the oxygen only slightly varies.
This is consistent with what expected from a typical CNO cycle (Clayton 1983).
The nova hypothesis first proposed for BY Cam (BM87) does not seem to be confirmed by the 
determination of a relatively low white dwarf temperature (Sion, these proceedings) 
and the efficiency of re-accretion is also questionable (Stehle \& Ritter 1999). However, the recent
discovery of an unexpected high H$_2$ column density in front of the source (Mouchet et al. 2003)
may be an indication of a significant circumstellar material related to nova activity.
The increasing number of sources now showing peculiar line ratios makes however this hypothesis not the 
unique explanation and alternate scenarios have to be found.

In a recent work, Schenker et al. (2002) have pointed out the importance of pre-evolution of CV systems
to explain the characteristics of the propeller system AE Aqr which is suggested to descend from a 
supersoft X-ray binaries. 
It is also worth to note that different other close binary systems such as the 
intermediate-mass X-ray binary Her X-1 (Jimenez-Garate 2002) and the low-mass soft X-ray transient
XTEJ1118+480 (Haswell et al. 2002) have now been found to show non-solar abundances 
with values N/C=14-20 and N/C$\geq$\,6 respectively.
In all cases, an evolution effect of the companion is invoked as the most probable origin. 
This may therefore be an indication of a more general process at work in the evolution of binary systems 
which have been neglected up to now.


\begin{references}
\reference Bonnet-Bidaud, J.M., Mouchet, M. 1987, A\&A,  188, 89 (BM87)
\reference Clayton, D. 1983, Principles of Stellar Evolution, Chicago Univ Press
\reference Ferland, G.J., Korista, K.T., Verner, D.A., et al. 1998, PASP, 110, 761
\reference Haswell, C., Hynes, R., King, A. et al. 2002, MNRAS 332, 928 
\reference Jimenez-Garate, M., Hailey, C., den Herder, J. et al. 2002, ApJ 578, 391
\reference Langer, S.H., Chanmugam, G., Shaviv, G. 1982, ApJ, 258, 289
\reference Mauche, C.W., Lee, Y.P., \& Kallman, T.R. 1997, ApJ,  477, 832
\reference Mouchet, M., Bonnet-Bidaud, J.M., Hameury, J.M. 1990, in 
"Accretion-Powered Compact Binaries", ed C. Mauche (CUP), 247
\reference Mouchet, M., Bonnet-Bidaud, J.M., Somov, N.N., Somova, T.A. 1997, A\&A, 324, 109
\reference Mouchet, M., Bonnet-Bidaud, J.M., Abada-Simon, M. et al. 2002, in "Classical Novae Explosions", 
AIP Conf Proc. 637, 67
\reference Mouchet, M., Bonnet-Bidaud, J.M., Roueff, E. et al. 2003, A\&A (in press)
\reference Ramsay, G., Mason, K., Cropper, M., et al. 1994, MNRAS, 270, 692
\reference Remillard, R.A., Bradt, H.V., McClintock, J.E., et al. 1986, ApJ, 302, L11
\reference Schenker, K., King, A.R., Kolb, U., Wynn, G.A., Zhang, Z. 2002, MNRAS, 337, 1105
\reference Schmidt, G.D., Stockman, H.S. 1991, ApJ, 371, 749
\reference Stockman, H.S., Schmidt, G.D. 1996, ApJ, 468, 883
\reference Schmidt, G.D., Stockman, H.S. 2001, ApJ, 548, 410
\reference Silber, A., Bradt, H.V., Ishida, M., et al. 1992, ApJ, 389, 704
\reference Stehle, R., Ritter, H. 1999, MNRAS, 309, 245 
\reference Szkody, P., Silber, A. 1996, AJ, 112, 239
\end{references}
\end{document}